\documentclass[conference]{IEEEtran}
\IEEEoverridecommandlockouts
\usepackage{cite}
\usepackage{textcomp}
\usepackage{xcolor}
\def\BibTeX{{\rm B\kern-.05em{\sc i\kern-.025em b}\kern-.08em
    T\kern-.1667em\lower.7ex\hbox{E}\kern-.125emX}}

\usepackage[utf8]{inputenc}
\usepackage[T1]{fontenc}
\usepackage{amsmath,amssymb,amsbsy}
\usepackage{mathtools}
\usepackage{xparse}

\usepackage{graphicx}
\usepackage{dcolumn}
\usepackage{bm}
\usepackage{lipsum}
\usepackage{amsmath}
\usepackage{amssymb}
\usepackage{float}
\usepackage{setspace}

\usepackage[normalem]{ulem}
\usepackage{color}
\usepackage{caption,subcaption}
\usepackage{comment}
\usepackage{array,multirow,booktabs}
\usepackage[makeroom]{cancel}
\usepackage{amsthm}
\usepackage{mathtools}
\usepackage{comment}
\usepackage{standalone}
\usepackage{multicol}

\usepackage{cprotect}

\usepackage[colorlinks=true]{hyperref}
\usepackage{doi}
\usepackage{braket}
\usepackage{breqn}

\usepackage{enumitem}

\theoremstyle{definition}

\newtheorem{observation}{Observation}

\newcommand{\poly}{\mathit{poly}}
\newcommand{\Id}{\mathit{Id}}

\newcommand{\eps}{\epsilon}

\NewDocumentCommand{\bbeta}{o}{
    \pmb{\beta}\IfValueT{#1}{_{[#1]}}
}
\NewDocumentCommand{\bgamma}{o}{
    \pmb{\gamma}\IfValueT{#1}{_{[#1]}}
}
\NewDocumentCommand{\bgstate}{o}{
    \ket{\IfValueTF{#1}{\bbeta[#1]}{\bbeta},
        \IfValueTF{#1}{\bgamma[#1]}{\bgamma}
    }
}
\NewDocumentCommand{\bgstateT}{o}{
    \bra{   \IfValueTF{#1}{\bbeta[#1]}{\bbeta},
        \IfValueTF{#1}{\bgamma[#1]}{\bgamma}
    }
}

\title{Threshold-Based Quantum Optimization%
\thanks{Research presented in this article was supported by the Laboratory Directed Research and Development program of Los Alamos National Laboratory under project numbers 20200671DI/20190495DR.
\hfill LA-UR-21-25835
}}
\author{
    \IEEEauthorblockN{John Golden\IEEEauthorrefmark{1}, 
    Andreas Bärtschi\IEEEauthorrefmark{1}, 
    Daniel O'Malley\IEEEauthorrefmark{2} and 
    Stephan Eidenbenz\IEEEauthorrefmark{1}}
    \IEEEauthorblockA{\IEEEauthorrefmark{1}
    \textit{CCS-3 Information Sciences, Los Alamos National Laboratory},
    Los Alamos, NM 87544, USA \\
    Email: golden@lanl.gov, baertschi@lanl.gov, eidenben@lanl.gov
    }
    \IEEEauthorblockA{\IEEEauthorrefmark{2}
    \textit{EES-16 Earth Sciences, Los Alamos National Laboratory},
    Los Alamos, NM 87544, USA \\
    Email: omalled@lanl.gov
    }
}

\begin{document}

\maketitle

\begin{abstract}
    We propose and study Th-QAOA (pronounced Threshold QAOA), a variation of the Quantum Alternating Operator Ansatz (QAOA) that replaces the standard phase separator operator, which encodes the objective function, with a threshold function that returns a value $1$ for solutions with an objective value above the threshold and a $0$ otherwise. We vary the threshold value to arrive at a quantum optimization algorithm. We focus on a combination with the Grover Mixer operator; the resulting GM-Th-QAOA can be viewed as a generalization of Grover's quantum search algorithm and its minimum/maximum finding cousin to \emph{approximate} optimization.

    Our main findings include: (i) we provide intuitive arguments and show empirically that the optimum parameter values of GM-Th-QAOA (angles and threshold value) can be found with $O(\log(p) \times \log M)$ iterations of the classical outer loop, where $p$ is the number of QAOA rounds and $M$ is an upper bound on the solution value (often the number of vertices or edges in an input graph), thus eliminating the notorious outer-loop parameter finding issue of other QAOA algorithms; (ii) GM-Th-QAOA can be simulated classically with little effort up to 100 qubits through a set of tricks that cut down memory requirements; (iii) somewhat surprisingly, GM-Th-QAOA outperforms non-thresholded GM-QAOA in terms of approximation ratios achieved. This third result holds across a range of optimization problems (MaxCut, Max k-VertexCover, Max k-DensestSubgraph, MaxBisection) and various experimental design parameters, such as different input edge densities and constraint sizes. 
\end{abstract}

\section{Introduction}
Using the power of quantum computing to solve combinatorial optimization problems, such as Minimum Traveling Salesperson, Maximum Satisfiability, or Maximum Cut, has been one of the main drivers of the development of quantum computing theory and practice. Early hopes (mid-1990s) of using quantum computers to solve NP-hard optimization problems in polynomial time had to be tempered after the discovery of oracles relative to which $NP$ is not contained in $BQP$~\cite{bennett1997npbqp}; thus, the existence of polynomial-time quantum algorithms for NP-complete problems is highly unlikely albeit not impossible. Polynomial factor speed-ups over the best known classical alternatives are an active area of quantum algorithms research that -- often as a side effect -- expose new facets of the boundary between $NP$ and $BQP$.

Defined for general gate-level quantum computing, the Quantum Approximate Optimization Algorithm (QAOA), \cite{Farhi2014}, which was later generalized to the Quantum Alternating Operator Ansatz (QAOA) \cite{hadfield_qaoa}, can be seen as low-order Trotterization of adiabatic computing with promising initial results in the form of provable approximation guarantees for MaxCut on 3-regular graphs~\cite{Farhi2014} and the E3Lin2 problem on bounded occurrence instances~\cite{Farhi2015}. Additional provable QAOA-based approximation ratio guarantees have been largely elusive until now, but the value of QAOA as an optimization heuristic is immense. Akin to classical optimization heuristics, properties other than approximation ratio guarantees need to be studied: circuit depth, convergence guarantees, average-case performance on random or constructed instances, outer loop parameter finding methods, etc. 

The seminal work of the Quantum Alternating Operator Ansatz (QAOA) \cite{hadfield_qaoa} defines a framework that allows modeling of almost any combinatorial optimization problem as a QAOA problem with the objective functions as the sum of Pauli operator-based Hamiltonians forming the phase separator operator and a few variations of mixing unitaries. Follow-on work has focused on ways of efficiently finding outer loop parameters, e.g., \cite{khairy2019reinforcement}, the efficient preparation of high-quality initial states such as Dicke states \cite{baertschi2019deterministic}, Grover-inspired mixer unitaries \cite{baertschi2020grover,akshay2020reachabilitydeficits}, experimental studies of more optimization problems \cite{cook2020kVC}, and studies on the XY-mixer model \cite{nasa2020XY}.

In this paper, we propose  Th-QAOA (pronounced Threshold QAOA) -- a variation of the Quantum Alternating Operator Ansatz (QAOA) that replaces the standard phase separator operator, which encodes the objective function, with a threshold function that returns a value $1$ for solutions with an objective value above the threshold and a $0$ otherwise (Section \ref{sec:thqaoa}). We vary the threshold value to arrive at a quantum approximate optimization algorithm. Th-QAOA can be combined with any of the previously studied Mixers for QAOA (e.g., Transverse Field based $X$-mixer~\cite{Farhi2014,Farhi2015} for unconstrained problems, $XY$-model Ring and Clique Mixers for Hamming weight-constrained problems~\cite{nasa2020XY,cook2020kVC}); in this paper we focus on the combination with Grover Mixers (which have been introduced for both unconstrained~\cite{akshay2020reachabilitydeficits} and constrained~\cite{baertschi2020grover} optimization problems), which we denote by GM-Th-QAOA.

We first show in Section \ref{sec:outer} through a mix of formal analysis and experimental evidence that the optimum values for of GM-Th-QAOA, namely the threshold value $th$ as well as the angles $\beta_i, \gamma_i$ for each QAOA round $i\leq p$ with $p$ being the total number of QAOA rounds, can be found with a simple algorithm in $O(\log p \times \log M)$ iterations of the classical outer loop, where $M$ is an upper bound on the solution value. In the graph or logical optimization problems that we consider, $M$ is usually of the same order as the input size, e.g.,  the number of vertices or edges in an input graph for problems such as MaximumCut.  Our parameter finding algorithm thus eliminates the notorious outer-loop parameter finding issue of other QAOA algorithms. Our algorithm relies on insights from numerical simulation experiments that show that as we increase the threshold value (under fixed round count), the best approximation ratio that GM-Th-QAOA finds first increases monotonically, reaching a peak and then decreasing. This stable shape allows us to find the peak value with the optimum threshold value via an exponential search. 

As a second contribution, we describe (Section \ref{sec:fastgrover})
a method of efficiently simulating GM-Th-QAOA circuits classically (after an initial precomputation of objective values) whenever we are only interested in calculating the expectation value of the objective function of the selected solution, which is all that we need to assess the performance of GM-Th-QAOA. Our method works particularly well when coupled with the Grover Mixer \cite{baertschi2020grover} and used on constrained optimization problems. Constrained optimization problems are problems where not all computational basis states correspond to feasible solutions, but the set of feasible solutions is constrained in some ways, such as only including solutions with exactly $k$ qubits set to 1. Examples of constrained problems are Max k-Densest Subgraph, which asks for a subset $V'$ of exactly $k$ of the vertices of an input graph such that a maximum number of edges have both end vertices in set $V'$, $k$-VertexCover, Max Bisection (which we define later); MaximumCut is an example of an unconstrained problem. Our simulation method limits the state vector calculation to the subspace of feasible solutions. In addition, leveraging a feature of the Grover Mixer that solutions of equal objective function value all are assigned the same amplitude, we can further reduce our search space to all possible objective function values, which is usually $M$ as we are dealing with discrete values (such as counting edges). 
Our simulation methods allow us to simulate high round counts nearly effortlessly (compared to the precomputation) and to go to large problem sizes (90+ vertices) with relative ease, giving us much greater confidence that we may indeed be observing performance trends that can be extrapolated to larger problem instances, particularly when compared to the standard problem size in classical QAOA simulations (in most other works, including our own), which is around graph size 10--16.

In our third contribution (Section \ref{sec:experiments}), we present a large study of experimental results comparing GM-Th-QAOA with a Grover Mixer QAOA using the standard objective-value based phase separator. Our experiments are for Maximum k-Densest Subgraph, Maximum k-Vertex Cover, Maximum Bisection, and Maximum Cut, with graph sizes ranging up to 90, three different edge probabilities (in a random graph model), and three different constraint values (parameter $k$) where applicable. Our main experimental finding initially surprised us:
Threshold QAOA with the Grover Mixer outperforms the traditional objective-value based QAOA with Grover Mixer across all optimization problems and across all parameter values that we tested. This experimental result, coupled with the fact that we can find optimum outer loop parameter values much faster than for standard QAOA, makes GM-Th-QAOA a strong contender for best-in-class quantum optimization heuristic. 
In addition, our experiments validate the scalability of our fast simulation method. Finally, we observe that increasing the number of rounds beyond polynomial in input size leads GM-Th-QAOA to degenerate into Grover search, which -- on the positive side -- guarantees Grover-style speed-up, and -- on the negative side -- makes it questionable that QAOA's worst-case running time guarantee can be better than Grover's, or in other words, QAOA maybe no more effective than Grover's unstructured search in the worst case. While we appear to see such exponential blow-up of round counts in a few examples, most real-life examples are far from worst-case, however. Thus we conclude that GM-Th-QAOA is a very promising quantum heuristic for finding high-quality approximate solutions for optimization problems.

\section{Threshold QAOA}
\label{sec:thqaoa}
\subsection{Definition}
Suppose we are given an instance $I$ of a combinatorial optimization problem over inputs $x \in S$, where $S \subseteq \{0,1\}^n$ is the set of all feasible solutions; let $C(x)\colon S \to \mathbb{R}$ be the objective function which evaluates the cost of solution $x$. W.l.o.g., we look at maximization problems (as opposed to minimization), where we want to find maximum or close-to-maximum value solutions. $I$ could be an instance of Maximum Satisfiability, defined on $n$ binary variables with, say $m$ clauses. If all variable allocations are feasible solutions (i.e., $S = \{0,1\}^n$), we say that the problem is unconstrained, such as Maximum Satisfiability; if $S \subsetneq \{0,1\}^n$, we call the problem constrained, such as Max $k$-Densest Subgraph, which asks for a set of $k$ vertices from a given graph with a maximum number of edges with edges in the induced subgraph. Let
\begin{equation}
  \delta(x) =
    \begin{cases}
      0 & \text{if } C(x) \leq  th\\
      1 & \text{otherwise}
    \end{cases}       
\end{equation}
be an indicator function that assigns a value of $1$ to all solutions with an objective function above threshold $th$ and zero otherwise. 
The quantum subroutine of Th-QAOA is defined on an input tuple $(I, th, \ket{\psi}, H_P, H_M, p,\vec{\beta}, \vec{\gamma} )$, where:

\begin{itemize}
    \item $\ket{\psi}$ is the initial state,
    \item $H_P \ket{x} = \delta(x) \ket{x}$ is a phase separator Hamiltonian,
    \item $H_M$ is a mixer Hamiltonian,
    \item$p$ is the number of rounds/levels to run the algorithm, and
    \item $\vec{\gamma} = (\gamma_1,...,\gamma_p)^T$ and $\vec{\beta} = (\beta_1,...,\beta_p)^T$, each of length $p$ two real vectors. 
        These values are often called angles.
\end{itemize}
The Th-QAOA algorithm returns a quantum state after preparing the initial state $\ket{\psi}$, and applying $p$ rounds of the alternating simulation of the phase separator Hamiltonian for time $\gamma_i$ and the mixer Hamiltonian for time $\beta_i$:
\begin{align}
    \hspace*{-6pt}\ket{\text{Th-QAOA}} = \underbrace{e^{-i\beta_p H_M} e^{-i\gamma_p H_P}}_{\text{round }p}\cdots \underbrace{e^{-i\beta_1 H_M} e^{-i\gamma_1 H_P}}_{\text{round }1} \ket{\psi} \label{eq:th-qaoa}
\end{align}
In each round, $H_P$ is applied first, which separates the basis states of the state vector by phases $e^{-i\gamma \delta(x)}$. The mixing operator $H_M$ then provides parameterized interference between solutions of different cost values. After $p$ rounds, the state $\ket{\text{Th-QAOA}}$ is measured in the computational basis and returns a sample solution $y$ of cost value $C(y)$ with probability $|\braket{y|\text{Th-QAOA}}|^2$. 
The only difference to standard QAOA is our definition of the phase separator $H_P$ through a threshold delta function $\delta(x)$ instead of the cost value $C(x)$. 
The threshold parameter $th$ should only take on values below a known upper bound of the combinatorial problem, e.g., $th \leq m$ in our MaxSat problem; for most problems (with the notable exception of number problems, such as Min Traveling Salesperson), $th$ can be bounded polynomially in the size of the input problem, which will allow us to search for optimum threshold values in polynomial or even logarithmic outer loops. 

Although we use $\delta(x)$ as the function in the phase separator, we still evaluate the quality of the expectation value over the original objective function $C$ as $\bra{\text{Th-QAOA}}C\ket{\text{Th-QAOA}}$. For a maximization problem, we say Th-QAOA achieves an approximation ratio $\frac{\bra{\text{Th-QAOA}}C\ket{\text{Th-QAOA}}}{\max_{x\in S} C(x)}$.

\subsection{Connection to Grover's algorithm} 
The QAOA framework, including Th-QAOA, can be applied to a wide variety of problems and problem-specific mixers~\cite{hadfield_qaoa}.
Here we use the Grover Mixer, which has been studied for both unconstrained~\cite{akshay2020reachabilitydeficits} and constrained problems~\cite{baertschi2020grover}, and can be used whenever there exists an efficient state preparation unitary $U_S$ that prepares the equal superposition of all feasible states $S$, $\ket{S} := |S|^{-1/2}\sum_{x \in S} \ket{x}$.
Then we have a Grover Mixer Hamiltonian $H_{GM} := \ket{S}\bra{S}$ with Grover Mixer~\cite{baertschi2020grover}
\begin{equation}
\begin{aligned}
    e^{-i\gamma H_{GM}} 
    &= \Id - (1 - e^{-i\beta}) \ket{S}\bra{S} \\
    &= U_S (\Id - (1-e^{-i\beta}\ket{0}\bra{0})U_S^{\dagger}. \label{eq:gm}
\end{aligned}
\end{equation}
We call this combination of a threshold-based phase separator with the Grover Mixer GM-Th-QAOA,
and obtain an algorithm that shares many traits with Grover's quantum search algorithm~\cite{G96} and both
variational versions thereof~\cite{grover2005fixed,yoder2014fixed,biamonte2018variational} and its minimum/maximum finding version~\cite{DH96}. 
We thus bring these purely search-oriented algorithms into the approximate optimization realm, similar to the original 
Quantum Approximate Optimization Algorithm which was inspired by the minimum/maximum finding adiabatic algorithm.
See the Related Works section~\ref{sec:related-work} for more details on this connection.

\subsection{Completing the Picture: Outer Loops}
For any practical use, the core quantum subroutine of the Threshold QAOA algorithm as described needs to be embedded in a classical outer loop algorithms to find good values for input parameters $th, \vec{\beta}$, and $\vec{\gamma}$. Quickly finding good values for these variational parameters has been a challenge and the focus of a sizable fraction of the QAOA literature,
see e.g.~\cite{Farhi2014,Farhi2015,wang2018fermionic,nasa2020XY,zhou2018quantum}.

For a reverse picture of the GM-Th-QAOA with an arbitrary but fixed threshold $th$ and a \emph{variational} number of rounds $p$ values, we provide intuition through an analysis of solution distributions and backed up by experiments that the optimum angle parameters $\vec{\beta}$, and $\vec{\gamma}$ can be set to $\beta_i= \gamma_i = \pi$ for all rounds $i<p$, and the values for $\beta_p$ and $\gamma_p$ can be found through a linear search. 

For a fixed number of rounds $p$ as is common in the QAOA framework, we can find the optimum threshold value $th^*$ for GM-Th-QAOA, 
which we define as the value that maximizes the achieved approximation ratio $r$, through a modified binary search because the approximation ratios monotonically increase before monotonically decreasing again with increasing threshold value resulting in a single peak (see Figure 1 for an example). Combining this with the observation above, we find that for small $p$ the optimal parameters will be $\beta_i = \gamma_i$ for all rounds $i$, while for larger $p$ we can again deploy binary search for a transition round $t$ such that $\beta_i = \gamma_i = \pi\ \forall i<t$ and 
$\beta_i = \gamma_i = 0\ \forall i>t$, with angles $\beta_t, \gamma_t$ that can be found with a fine grid search.

The ease of finding these values in an outer loop is a key advantage of GM-Th-QAOA over other variations. 
The total running time of GM-Th-QAOA is thus the product of 
\begin{itemize}
    \item   the quantum subroutine running time $O(p(T_{GM}+t_{PS}))$, where $t_{GM},t_{PS}\in \poly(M)$ are the circuit depths of the Grover mixer and phase separator unitaries,
    \item   $\poly(M)$ iterations thereof (for fixed parameters $th,\beta,\gamma$) to reliably estimate  $\bra{\text{GM-Th-QAOA}}C\ket{\text{GM-Th-QAOA}}$, where the exact polynomial depends on the concentration of sample distribution 
    from $\ket{\text{GM-Th-QAOA}}$~\cite{Farhi2014,cook2020kVC},
    \item   $O(\log p)$ time to find the transition round $t$ and angles $\beta_t, \gamma_t$ for a candidate threshold $th$,
    \item   $O(\log M)$ time to find the optimum threshold value $th^*$.
\end{itemize}
We show in Section~\ref{sec:fastgrover} that after an initial precomputation of the objective values of all feasible states $S$, 
the first two points can be classically simulated efficiently.


\section{Finding high-quality parameter values efficiently for GM-Th-QAOA}
\label{sec:outer}
We use a structural characterization of the distribution of objective values among feasible solutions to build a simple heuristic for finding high-quality angle values for GM-Th-QAOA. 
For ease of presentation, we adopt the convention that the GM-Th-QAOA phase separator only acts on states strictly greater the threshold, rather than greater than or equal to the threshold. Therefore, if our optimization instance $I$ has a best solution $x^*$ with objective value $C(x^*)$, the threshold that will select states of this score is $C(x^*) -1$. 
which we will call the  \textbf{maximal threshold}.
Moreover, as the number of rounds decreases, the threshold which returns the highest approximation ratio also decreases. Therefore, the maximal threshold is often not the threshold that returns the highest approximation ratio. 
We thus call the threshold value $th^*$, which returns the highest approximation ratio for a given number of rounds $p$ the \textbf{optimal threshold}.

We begin with some notation: after $p$ rounds of Th-QAOA with threshold $th$, we have the state
\begin{equation}
\begin{aligned}
    \ket{S^{(p)}}   =&\ c_0^{(p)}{\left(\sum\text{states with score}\le th\right)}+ \\
                    &\  c_1^{(p)}{\left(\sum\text{states with score}>th\right)}
\end{aligned}
\end{equation}
with coefficients at round 0 as $c_0^{(0)} = c_1^{(0)} := |S|^{-1/2}$ for $|S|$ feasible solutions.
For a larger number of rounds $p$, we have the recursive form:
\begin{equation}\label{eq:var-grover-coeffs}
\begin{aligned}
    c_0^{(p)}   =&\ c^{(p-1)}_0 - \left(1-e^{-i\beta}\right)\left(r c^{(p-1)}_0 + (1-r)c^{(p-1)}_1 e^{-i\gamma}\right),\\
    c_1^{(p)}   =&\ c^{(p-1)}_1 e^{-i \gamma} 
                    -\left(1-e^{-i\beta}\right)\left(rc^{(p-1)}_0 + (1-r)c^{(p-1)}_1 e^{-i\gamma}\right).
\end{aligned}
\end{equation}
Here, we have used the basic definitions of the Th-QAOA algorithm~\eqref{eq:th-qaoa} and the Grover Mixer~\eqref{eq:gm} and dropped indices $p-1$ for the angle labels to ease notation. Label $r$ denotes the fraction of states with score $\le th$. 
These states have degeneracy $d_0$ and $d_1$, respectively, and we have $d_0 + d_1 = |S|$.
Thus, using $r \equiv \frac{d_0}{|S|}$,
we arrive at Eq.~(\ref{eq:var-grover-coeffs}).

\subsection{First-round Angle Values}
Let us start by exploring the first round in detail.
We have 
\begin{equation}
\begin{aligned}
    c_0^{(1)}   =& |S|^{-1/2}\left(1 - \left(1-e^{-i\beta}\right)\left(r + (1-r)e^{-i\gamma}\right)\right),\\
    c_1^{(1)}   =& |S|^{-1/2}\left(e^{-i\gamma} -\left(1-e^{-i\beta}\right)\left(r + (1-r)e^{-i\gamma}\right)\right),\\
\end{aligned}
\end{equation}
Taking the absolute value squared of these coefficients gives
\begin{align}
\begin{aligned}
    \left\|c_0^{(1)}\right\|^2 = |S|(&1+2 (r-1) (\sin (\beta ) \sin (\gamma )+ \\
    &4 (2 r-1) \sin ^2(\beta/2)\sin ^2(\gamma/2))),
\end{aligned}\\
\begin{aligned}
    \left\|c_1^{(1)}\right\|^2 = |S|(&1+2 r (\sin (\beta ) \sin (\gamma )+ \\ &4 (2 r-1) \sin ^2(\beta/2)\sin ^2(\gamma/2)))
\end{aligned}
\end{align}

\begin{observation} 
    When $r<\frac{3}{4}$, 
    \begin{equation}
        \beta = \gamma = \arctan\left(-\sqrt{3-4 r},1-2r\right)
    \end{equation}
    gives $\left\|c_0^{(1)}\right\|^2=0$, where $\arctan(x,y)$ takes into account which quadrant the point $(y,x)$ is in when calculating $\arctan(x/y)$.
\end{observation}

\begin{observation}
For $r\ge \frac{3}{4}$, $\beta = \gamma = \pi$ minimizes $\left\|c_0^{(1)}\right\|^2$ and maximizes $\left\|c_1^{(1)}\right\|^2$ and $\braket{\psi^{(1)}|H_P|\psi^{(1)}}$.
\end{observation}

In other words, when $r<\frac{3}{4}$ there exist $\beta, \gamma$ that will entirely kill off all states with score $\le th$, and no further rounds are necessary. 
If $r\ge\frac{3}{4}$, setting $\beta=\gamma=\pi$ gives the highest approximation ratio possible after round 1, and further rounds are necessary to achieve $\left\|c_0\right\|^2=0$.

Observation 1 can be verified by explicit calculation.
For Observation 2, we give the following proof.
\begin{proof}
Our goal is to maximize
\begin{align}
\begin{aligned}
    \left\|c_1^{(1)}\right\|^2 = |S|(&1+2 r (\sin (\beta ) \sin (\gamma )+ \\&
    4 (2 r-1) \sin ^2(\beta/2)\sin ^2(\gamma/2)))
\end{aligned}
\end{align}
for $r\ge3/4$. We re-write $r=3/4+\epsilon/8$ for $\epsilon\ge0$, and drop constant terms and overall factors to get the function we are trying to maximize:
\begin{equation}
    \sin (\beta ) \sin (\gamma )+(2 +\eps) \sin ^2(\beta/2)\sin ^2(\gamma/2).
\end{equation}
Our claim is that $\beta = \gamma = \pi$ are optimum values, i.e. our goal is now to prove
\begin{equation}
    \sin (\beta ) \sin (\gamma )+ (2+\eps) \sin ^2(\beta/2)\sin ^2(\gamma/2) \le 2+\eps,
\end{equation}
or, re-arranging terms,
\begin{equation}
\begin{aligned}
    \label{eq:bound}
    \sin ^2(\beta/2) \sin ^2(\gamma/2)+\frac{1}{2}\sin (\beta ) \sin (\gamma )\leq  \\
    1+\frac{\epsilon}{2}  \left(1-\sin ^2(\beta/2) \sin ^2(\gamma/2)\right).
\end{aligned}
\end{equation}
Consecutively using on the left side the trigonometric identities $\sin(2\theta) = 2\sin(\theta)\cos(\theta)$ and $\cos(\alpha\pm\beta) = \cos(\alpha)\cos(\beta)\mp\sin(\alpha)\sin(\beta)$, 
we bring Eq.~(\ref{eq:bound}) into the equivalent form
\begin{equation}
\begin{aligned}
    \cos(\beta/2-\gamma/2)^2-\cos ^2(\beta/2) \cos ^2(\gamma/2) \leq \\
    1+\frac{\epsilon}{2}  \left(1-\sin ^2(\beta/2) \sin ^2(\gamma/2)\right),
\end{aligned}
\end{equation}
which holds as $0\le\frac{\epsilon}{2}  \left(1-\sin ^2(\beta/2) \sin ^2(\gamma/2)\right)$ as well as 
$0\le \cos(\beta/2-\gamma/2)^2 \le 1$; $0\le\cos ^2(\beta/2) \cos ^2(\gamma/2) \le 1$.
\end{proof}

\subsection{Extending Beyond the First Round}
Based on numerical observations, we find
that the optimal angles for multi-round GM-Th-QAOA are straightforward. Specifically, we present a simple scheme to determine the minimum number of rounds $p$ necessary to achieve the maximum approximation ratio for a problem instance $I$ and threshold $th$ (this is akin to Grover's overshooting problem). For the first $p-1$ rounds, setting $\beta=\gamma=\pi$ gives optimal results, and we present an analytic formula for the optimal final angles $\beta_p,\gamma_p$. 

We begin by describing the optimal angles for round $p$, followed by the method of determining the ideal number of rounds $p$. For the first $p-1$ rounds, we set $\beta=\gamma=\pi$ and denote the coefficients with the notation $c^{(j)}_{i,\pi}$ ($j \le p-1, i\in\{0,1\}$). The coefficients are then given by
\begin{align}\label{eq:var-grover-pi-coeffs}
    &c_{i,\pi}^{(j<p)} = (-1)^i c^{(j-1)}_{i,\pi} - 2\left(r c^{(j-1)}_{0,\pi} - (1-r)c^{(j-1)}_{1,\pi}\right),\\
    &c_0^{(p)} = c_{0,\pi}^{(p{-}1)}-\left(1-e^{-i \beta }\right) \left(r c_{0,\pi}^{(p{-}1)}+(1-r)c_{1,\pi}^{(p{-}1)}e^{-i \gamma }\right).\nonumber
\end{align} 
Using the fact that $c_{i,\pi}^{(j<p)} \in \mathbb{R}$, we can solve $||c_0^{(p)}||^2 = 0$ to find the optimal angles
\begin{equation}
\begin{aligned} 
\label{eq:var-grover-angles}
\beta_p &= \arctan\left(-\Delta\left|c_{0,\pi}^{(p{-}1)}\right| ,\frac{2 (1-r)}{|S|}-\left(c_{0,\pi}^{(p{-}1)}\right)^2\right),\\
\gamma_p &= \arctan\left(-\frac{\Delta}{c_{1,\pi}^{(p-1)} \text{sgn}(c_{0,\pi}^{(p{-}1)})}, \frac{c_{0,\pi}^{(p{-}1)} (1-2 r)}{c_{1,\pi}^{(p-1)}}\right),
\end{aligned}
\end{equation}
with
\begin{equation}
    \Delta =  \sqrt{4 (1-r)|S|^{-1}-\left(c_{0,\pi}^{(p{-}1)}\right)^2}.
\end{equation}
Our main insight is now that Eq.~(\ref{eq:var-grover-angles}) only gives $\beta_p, \gamma_p \in \mathbb{R}$ when $\Delta \in \mathbb{R}$, i.e. 
\begin{equation}\label{eq:var-grover-rounds}
    4(1-r)-|S|\left(c_{0,\pi}^{(p{-}1)}\right)^2>0.
\end{equation}
If this inequality is not satisfied, then we have numerically observed 
that $\beta=\gamma=\pi$ again gives the best possible approximation ratio at round $p$, and further rounds are necessary to reach $||c_0||^2=0$. 

The inequality in Eq.~(\ref{eq:var-grover-rounds}) does not actually have any dependence on $|S|$, as Eq.~(\ref{eq:var-grover-pi-coeffs}) shows that $c_{0,\pi}^{(j<p)} = |S|^{-1/2}\times(\text{a polynomial in }r)$.
Therefore, for a given $p$, we can explicitly evaluate the recursion in Eq.~(\ref{eq:var-grover-pi-coeffs}) and determine the range of $r$ values for which Eq.~(\ref{eq:var-grover-rounds}) is satisfied, thus determining the number of rounds necessary to get $||c_0^{(p)}||^2=0$ as a function of $r$:
\begin{itemize}
    \item 1 round: $0\le r <3/4$
    \item 2 rounds: $3/4 \le r < \frac{1}{8} \left(\sqrt{5}+5\right)$
    \item 3 rounds: $ \frac{1}{8} \left(\sqrt{5}+5\right) \le r < 0.950484\ldots$
    \item 4 rounds:  $0.950484\ldots \le r < 0.969846\ldots$
\end{itemize}
We remark that the above is not a strict induction proof, as one could imagine a scenario where choosing a non-$\pi$ set of angles at some round $j$ allows for a greater approximation ratio (i.e. smaller $||c_0^{(p)}||^2$) to be reached in round $p$. 
However, in practice, this does not ever seem to be the case, as confirmed with a large number of numerical simulations.

\subsection{Completing the Picture: Angle Finding}\label{sec:var-grover-angles-algorithm}

Recall that an input tuple for the quantum subroutine of GM-Th-QAOA is $(I, th, \ket{\psi}, H_P, H_{GM}, p,\vec{\beta}, \vec{\gamma} )$. 
If we are given only a candidate threshold $th$, we find close-to-optimum angles $\vec{\beta}, \vec{\gamma}$ in $O(\log p)$ steps:
The idea is to find a transition round $t$ such that we have angles $\beta_i=\gamma_i = \pi$ for rounds $i<t$ and $\beta_i=\gamma_i = 0$ for rounds $t < i \leq p$:

\begin{enumerate}
    \item   Exponential search over the number of rounds $p_k = \lceil\lambda^k\rceil$ for small $\lambda>1$, 
            always using angles $\beta=\gamma=\pi$, until we find two values $k+1,k+2$ for which the $p_{k+1}$ rounds give a 
            higher expectation value than $p_{k+2}$ rounds.
    \item   Do a modified binary search over the number of rounds $p'$ in the interval $[p_k,p_{k+2}]$,
            to find maximum expectation value of a $p'$-round QAOA with angles $\beta=\gamma=\pi$.
    \item   The found $p'$ could be an overshoot or an undershoot of the optimum number of rounds, and tightly bounds the range of possible values for $r$.
            Hence we test both a $t=p'$-round and a $t=p'+1$-round QAOA with a linear search in $r$, with angles given by Eq.~(\ref{eq:var-grover-angles}).

\end{enumerate}
Our algorithm bears resemblance to the exponential quantum search algorithm~\cite{boyer1998} generalizing Grover's search to an unknown number of marked items.
Depending on the number of rounds and desired accuracy, the final step may not be necessary, as setting $\beta=\gamma=\pi$ for all rounds gives near-optimal results.
For small $p$ it will also be more efficient to do a linear search over the number of rounds.

\subsection{Completing the Picture: Threshold Finding}
\label{sec:var-grover-threshold-algorithm}
If we are given only $(I, \ket{\psi}, H_P, H_{GM}, p)$, we find the optimum threshold value $th^*$ by doing a modified binary search on threshold values $th$ with $0 \le th \leq M$, where M is the maximum possible threshold value. For each value $th$ that we test, we run the angle-finding algorithm from Section~\ref{sec:var-grover-angles-algorithm}. 

\begin{figure*}
\centering
\begin{subfigure}{.41\textwidth}
  \centering
  \includegraphics[width=\linewidth]{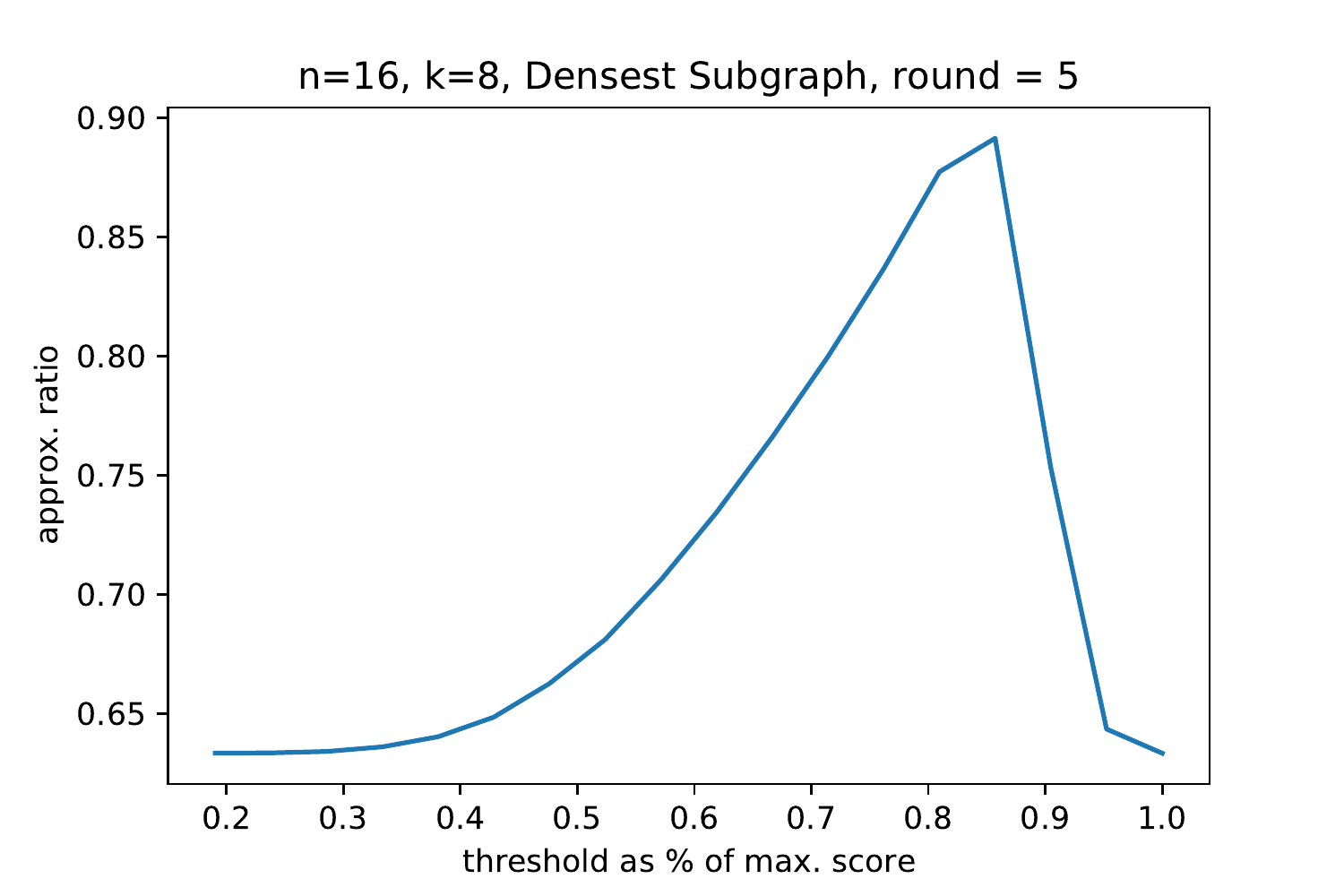}
\end{subfigure}
\hspace{0.05\linewidth}
\begin{subfigure}{.41\textwidth}
  \centering
  \includegraphics[width=\linewidth]{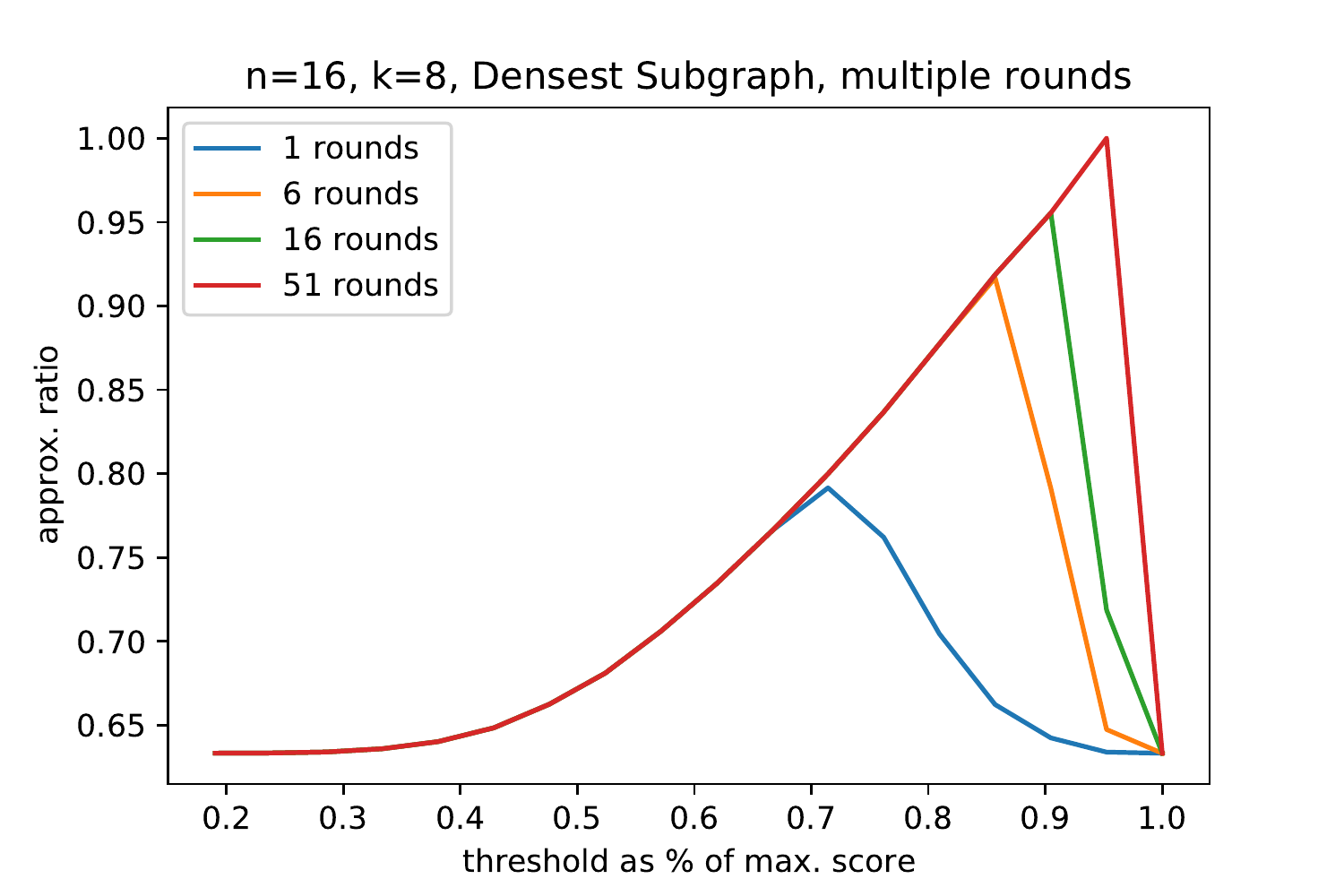}
\end{subfigure}
\caption{Approximation ratio as a function of the threshold $th$ (fraction of max score) for GM-Th-QAOA for $k$-Densest Subgraph: (left) shows the results for just a 5-round GM-Th-QAOA, (right) shows the evolution over many rounds for the same problem. This shows that at a given number of rounds $p$, the optimal threshold value increases monotonically up to a peak value and then decreases sharply. Furthermore, the threshold which produces the peak value for a given round $p$ increases as $p$ increases.}
\label{fig:threshold}
\end{figure*}

This simple search technique is based on the results of extensive numerical experiments, which showed that for a given number of rounds $p$ the approximation ratio as a function of $th$ increases monotonically up to a peak value and then monotonically decreases. 
Should such a single-peak observation ever fail, we can always perform a linear search instead, increasing the number of steps 
from $O(\log M)$ to $O(M)$.
Fig.~\ref{fig:threshold} (left) shows the single peak for a random problem over $5$ rounds. 

We again observe a resemblance to a quantum search algorithm: The maximum finding algorithm~\cite{DH96} 
keeps increasing a threshold $th$ which it passes to the mentioned exponential quantum search algorithm~\cite{boyer1998} to sample a new, higher threshold. This technique finds a maximum with a quadratic Grover speed-up over classical brute-force search.
Our algorithm behaves somewhat similarly in case the number of rounds $p$ is large enough to accommodate a meaningful use of large thresholds, see Fig.~\ref{fig:threshold} (right).
Furthermore, our approach allows for a potential increase in $th$ round-over-round, which we leave to a future work.

In terms of actual implementation on all our algorithms, we can always get the full statevector as long as we just simulate classically; however, on actual quantum computers with measurements and returns of single basis states only, we need to execute enough rounds to arrive at a good approximation of the expectation value. Standard statistics using Chebyshev's inequality tells us that $O(M^3)$ shots are sufficient to approximate the expectation value within an additive error of 1 (see, e.g., \cite{cook2020kVC} for a derivation in the QAOA context).

\section{Efficient Classical Numerical Evaluation of Grover Mixer-based Circuits}
\label{sec:fastgrover}
In preparation for our numerical experiments, we describe an efficient way to simulate the Grover Mixer QAOA for standard phase separators~\cite{baertschi2020grover,baertschi2021grover} and the Threshold-based GM-Th-QAOA. 
The main source of our relative scalability relies on leveraging the Grover Mixer feature that solutions which evaluate to the same objective value also have identical amplitudes (thereby also discarding all infeasible basis states). Our approach calculates the expectation value only, as opposed to computing full statevector amplitudes, but this is sufficient.

\begin{figure*}
\centering
\begin{align}
    U_M(\beta_p)e^{-i\gamma_p H_P}\ket{S^{(p-1)}} &= \left(\Id-\left(1-e^{-i\beta_p}\right)\ket{D^n_k}\bra{D^n_k}\right)\smash{\left(\sum_i c^{(p{-}1)}_i e^{-i\gamma_p h_i} \ket{d_i}\right)} \notag \\
    &=\sum_i\left[c^{(p{-}1)}_i e^{-i\gamma_p h_i} - \frac{1}{\binom{n}{k}} \left(1-e^{-i\beta_p}\right)\left(\sum_j c^{(p{-}1)}_j e^{-i\gamma_p h_j}\right)\right]\ket{d_i}.\label{eq:full-coeff}
\end{align}
\caption{Explicit derivation of coefficients for fast numerical evaluation}
\label{bigequation}
\end{figure*}

We recall the definition~\eqref{eq:gm} of the Grover Mixer~\cite{baertschi2020grover}, $e^{-i\beta \ket{S}\bra{S}}$, which requires
an efficient state preparation unitary preparing all feasible solutions $S$ in equal superposition: $\ket{S} := |S|^{-1/2}\sum_{x\in S} \ket{x}$.
For most of our experiments, this starting state is the Dicke state
\begin{align}
\ket{D^n_k} := \binom{n}{k}^{-1/2} \sum\nolimits_{HW(d_i)=k} d_i,
\end{align}
which is the equal superposition of all computational basis states $d_i$ on $n$ qubits with exactly $k$ one values, or more formally of Hamming weight $k$.  Quantum circuits that prepare $\ket{D^n_k}$ deterministically in depth $O(n)$ exist~\cite{baertschi2019deterministic}. 
After executing $p-1$ rounds of QAOA, we get the state
\begin{equation}
    \ket{S^{(p-1)}} := \sum c^{(p{-}1)}_i \ket{d_i},
\end{equation}
where the sum is over all the individual states appearing in the Dicke state. 
For the $p$-th round, we begin by applying the phase separator:
\begin{align}
    e^{-i\gamma_p H_P}\ket{S^{(p-1)}} = \sum_i c^{(p{-}1)}_i e^{-i\gamma_p h_i} \ket{d_i},  
\end{align}
where $H_P\ket{d_i} = h_i\ket{d_i}$.
We then then apply the Grover Mixer
\begin{equation}
    U_M(\beta_p) = I-\left(1-e^{-i\beta_p}\right)\ket{D^n_k}\bra{D^n_k},
\end{equation}
to our state $e^{-i\gamma_p H_P}\ket{S^{(p-1)}}$, yielding Eq.~\ref{eq:full-coeff} in Fig. \ref{bigequation}.
%
Beginning with $c_i^{(0)} = \binom{n}{k}^{-1/2}$,
we can thus directly calculate the coefficients of the $p$-th round straightforwardly in terms of the coefficients of the $(p-1)$-th round.
However, it is unnecessary to calculate the coefficient for every computational basis state, as the Grover Mixer ensures that all states with the same objective function value have the same coefficient. 
In other words, if $h_i = h_j$, then $c_i = c_j$ for all rounds. 
Therefore, once we have precomputed the set of possible values for $H_P\ket{d_i}$ and their degeneracies (i.e., how many states share the same objective value), we can significantly reduce the total number of calculations necessary as follows: let us assume that the set of all $\{h_i\}$ can be reduced to the set of $l$ distinct values $\{g_i\}$ with degeneracies $\{d_i\}$.
In many cases, the number of distinct objective function values is much smaller than the number of feasible solutions, $l\ll \binom{n}{k}$.
For example, with $k$ Densest Subgraph, $l$ is $\mathcal{O}(n^2)$.
If we are only interested in the Hamiltonian expectation value, $\braket{S^{(p)}|H_P|S^{(p)}}$, we need to only keep track of $l$ coefficients, reducing our computational load considerably. 
If we label our reduced set of $l$ coefficients $\{\hat{c}_i^{(p)}\}$, i.e. $\hat{c}_i^{(p)}$ is the coefficient for all states with objective cost $g_i$, we get from Eq.~(\ref{eq:full-coeff}) to 
\begin{align}
    \hat{c}_i^{(p)} = \hat{c}^{(p{-}1)}_i e^{-i\gamma_p g_i} - \frac{\left(1-e^{-i\beta_p}\right)}{\binom{n}{k}} 
    & \sum_{j=1}^l d_i\hat{c}^{(p{-}1)}_j e^{-i\gamma_p g_j}, \notag \\
    \label{eq:grover-fast-exp}
    \Rightarrow \braket{S^{(p)}|H_P|S^{(p)}} = 
    & \sum_{i=1}^l d_i g_i\left\|\hat{c}_i^{(p)}\right\|^2.
\end{align}

\begin{table*}[ht!]
    \centering
    \renewcommand{\arraystretch}{1.2}
    \begin{tabular}{@{}p{0.37\linewidth}r@{}c@{}p{0.19\linewidth}p{0.15\linewidth}@{}}
        \toprule
        \bf{Parameters}       & \bf{Optimization Problems} & \bf{\ and\ } & \bf{Grover Mixers}      & \bf{Phase Separators}   \\
        \midrule
        $n = 15, 20, 35, 40, 50, 60, 70, 80, 90, 100$    & k-Densest Subgraph & --- & $\ket{D^n_k}\bra{D^n_k}$        & Threshold                        \\
        $k = .25n, .5n,.75n, n-10$            & k-Vertex Cover & --- & $\ket{D^n_k}\bra{D^n_k}$            & Standard       \\
        Edge Probabilities $= .25, .5, .75$ & Max Cut & --- & $\ket{+^n}\bra{+^n}$                          &                   \\
        Number of rounds p: up to 16k & Max Bisection& --- & $\ket{D^n_{n/2}}\bra{D^n_{n/2}}$ & \\
        \bottomrule
    \end{tabular}
    \caption{Experimental Design: We performed numerical experiments with four different optimization problems and their corresponding Grover Mixers, comparing the standard and threshold phase separators. All input graphs were random graphs with edge probabilities from three different values. We studied graphs of up to 90 nodes, with $k$-values that define feasible solutions varied as three different fractions of node counts, as well as a standard $k = n-10$ case. We generated 30 random graphs for each combination.}
    \label{choices}
\end{table*}

\section{Experimental Evaluation}
\label{sec:experiments}

To test the performance of GM-Th-QAOA and compare it against standard GM-QAOA with regular phase separators, we execute a large set of experiments. Table \ref{choices} gives an overview of our experimental design. Our experiments were executed on a mid-range 64 core machine and took a few weeks; our software stack relies on standard numerical libraries using SciPy, NumPy in Python, and Julia. For each graph parameter combination (that is, the vertex count $n$ and edge probabilities), we generate 30 random graphs with corresponding edge probabilities. We executed all our runs with the different numbers of rounds $p$ by selecting values up to 16384. As phase separators, we used the threshold version, which is the key building block of GM-Th-QAOA and the standard Hamiltonian phase separator formulation. 

For each run of GM-Th-QAOA, we calculate the optimal outer loop parameters $th$, $\beta, \gamma$ as described in Section \ref{sec:outer} through a set of test runs. For the standard phase separator GM-QAOA runs, we use the basin hopping technique to find good $\beta, \gamma$ values, which is standard for QAOA simulations, while using similar precomputation and fast computation tricks as those described in Section~\ref{sec:fastgrover}. We set most basin hopping parameters to default values and observed that basin hopping converges to its optimum. For larger instances (i.e., $n>20)$, basin hopping becomes prohibitive in computational cost. Thus, we could not run all standard phase separator cases for high $n$ and $p$ counts, as evidenced in the plots. As other authors have observed \cite{cook2020kVC},\cite{nasa2020XY}, our confidence that basin hopping finds near-optimum angle values is based among other factors on the fact that for a small number of rounds (<6) with standard phase separators, basin hopping matches or beats the approximation ratios found through an exhaustive fine grid search. Thus, while it is possible that the truly best angles 
were not found by basin hopping for a few cases, our conclusions from the relative ranking of the two methods are sound.

We studied the following four graph optimization problems for input graphs $G = (V,E)$ with $n:=|V|$ vertices and $m:=|E|$ edges. They are all $NP$-hard, and have straightforward representations as Hamiltonians, see e.g., \cite{hadfield_qaoa}:
\paragraph{Max $k$-Densest Subgraph} Given an additional input parameter $k<n$, find a subset $V' \subset V$ of $k$ vertices with a maximum number of edges between vertices from $V'$.  
\paragraph{Max $k$-Vertex Cover} Given an additional input parameter $k<n$, find a subset $V' \subset V$ of $k$ vertices with a maximum number of edges with at least one end point in $V'$.  
\paragraph{Max Cut} Find a subset $V' \subset V$ of vertices with a maximum number of edges with one endpoint in $V'$ and the other in $V\setminus V'$. This is an unconstrained problem, thus every computational basis state represents a feasible solution.
\paragraph{Max Bisection} Find a subset $V' \subset V$ of $n/2$ vertices with a maximum number of edges with one end point in $V'$ and the other in $V\setminus V'$.  

Finally, we note that in the following comparisons we compare performance of GM-Th-QAOA and GM-QAOA given equivalent number of rounds $p$.
This should not necessarily be viewed as comparison given equivalent quantum resources, as we have not given an explicit construction of the threshold operator. 
We leave a more detailed discussion of the computational cost of these approaches, as well as other mixers, to a future work.

\subsection{GM-Th-QAOA Outperforms Standard GM-QAOA}

\begin{figure*}
\centering
\includegraphics[width=0.85\linewidth]{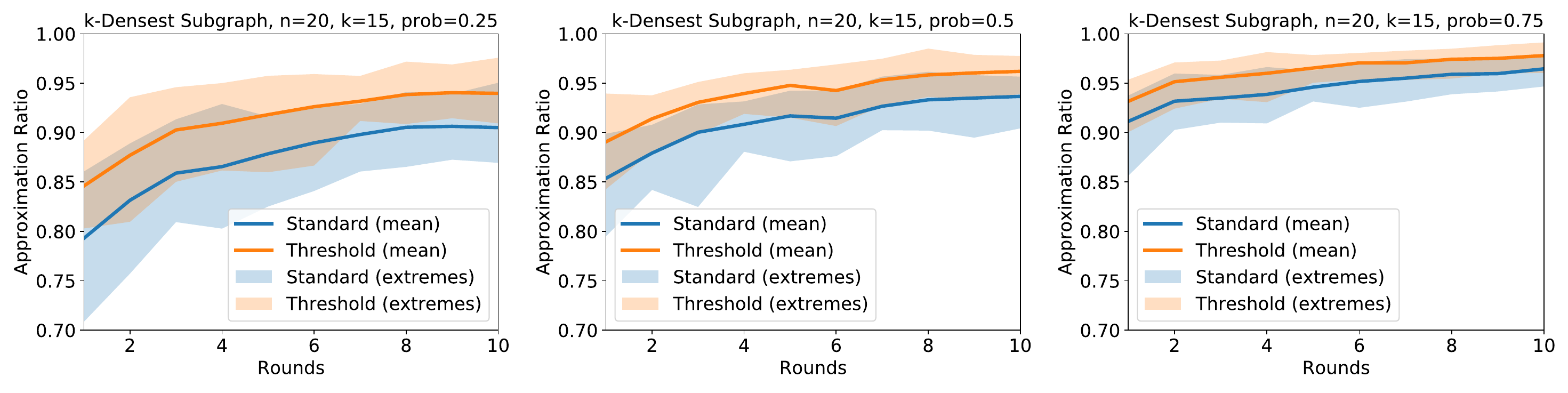}
\caption{{\bf Maximum k-Densest Subgraph}: QAOA performance for different edge probabilities}
\label{fig:kds-probs}
\end{figure*}

\begin{figure*}
\centering
\includegraphics[width=0.85\linewidth]{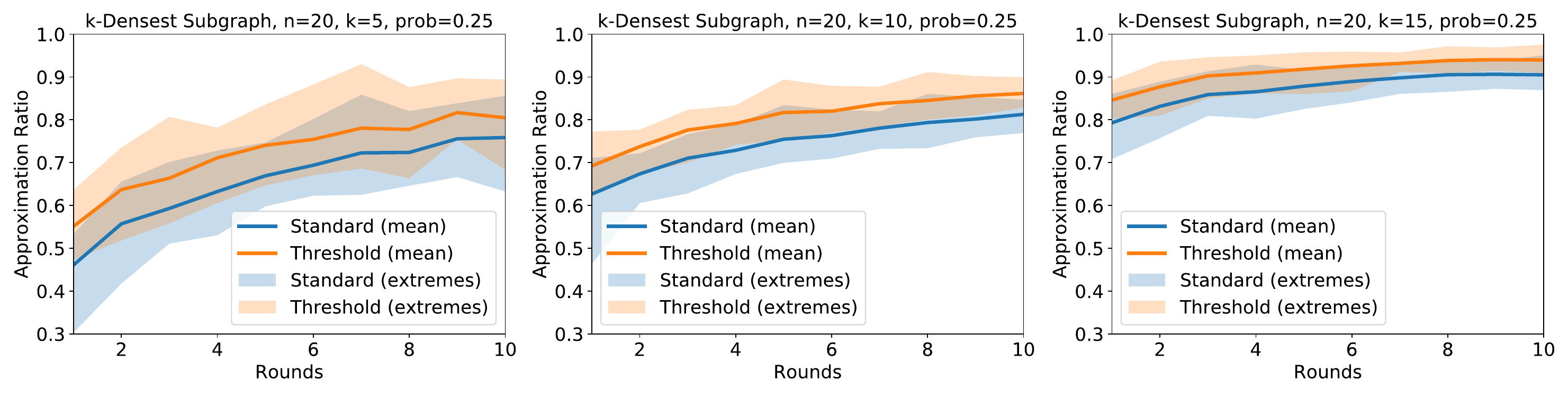}
\caption{{\bf Maximum k-Densest Subgraph}: QAOA performance for different $k$-values}
\label{fig:kds-ks}
\end{figure*}

\begin{figure*}
\centering
\includegraphics[width=0.85\linewidth]{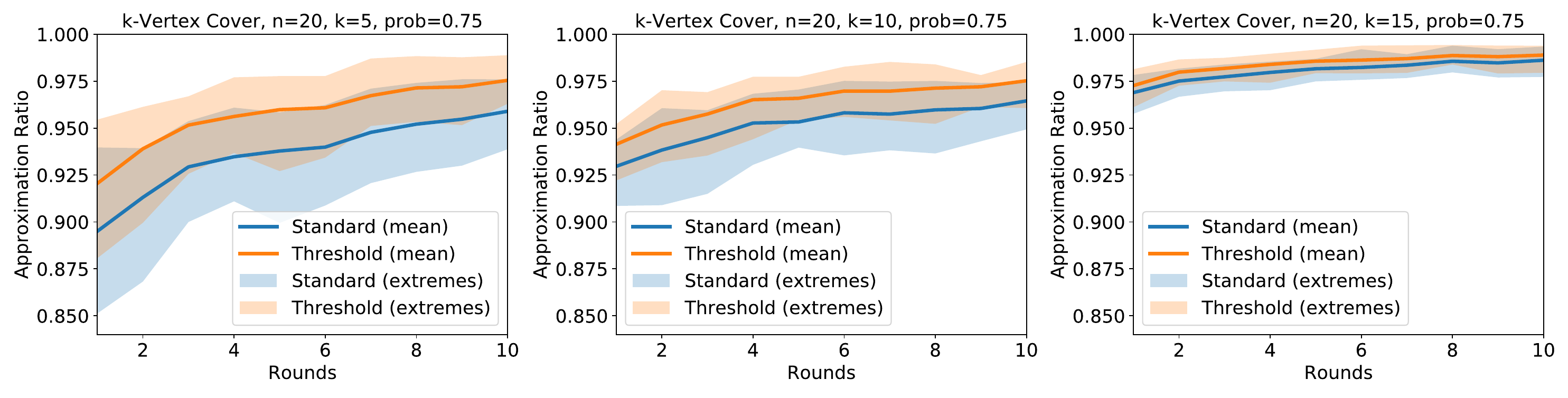}
\caption{{\bf Maximum k-Vertex Cover}: QAOA performance for different $k$-values}
\label{fig:kvc-ks}
\end{figure*}

\begin{figure*}
\centering
\hspace{1.5cm}
\includegraphics[width=0.9\linewidth]{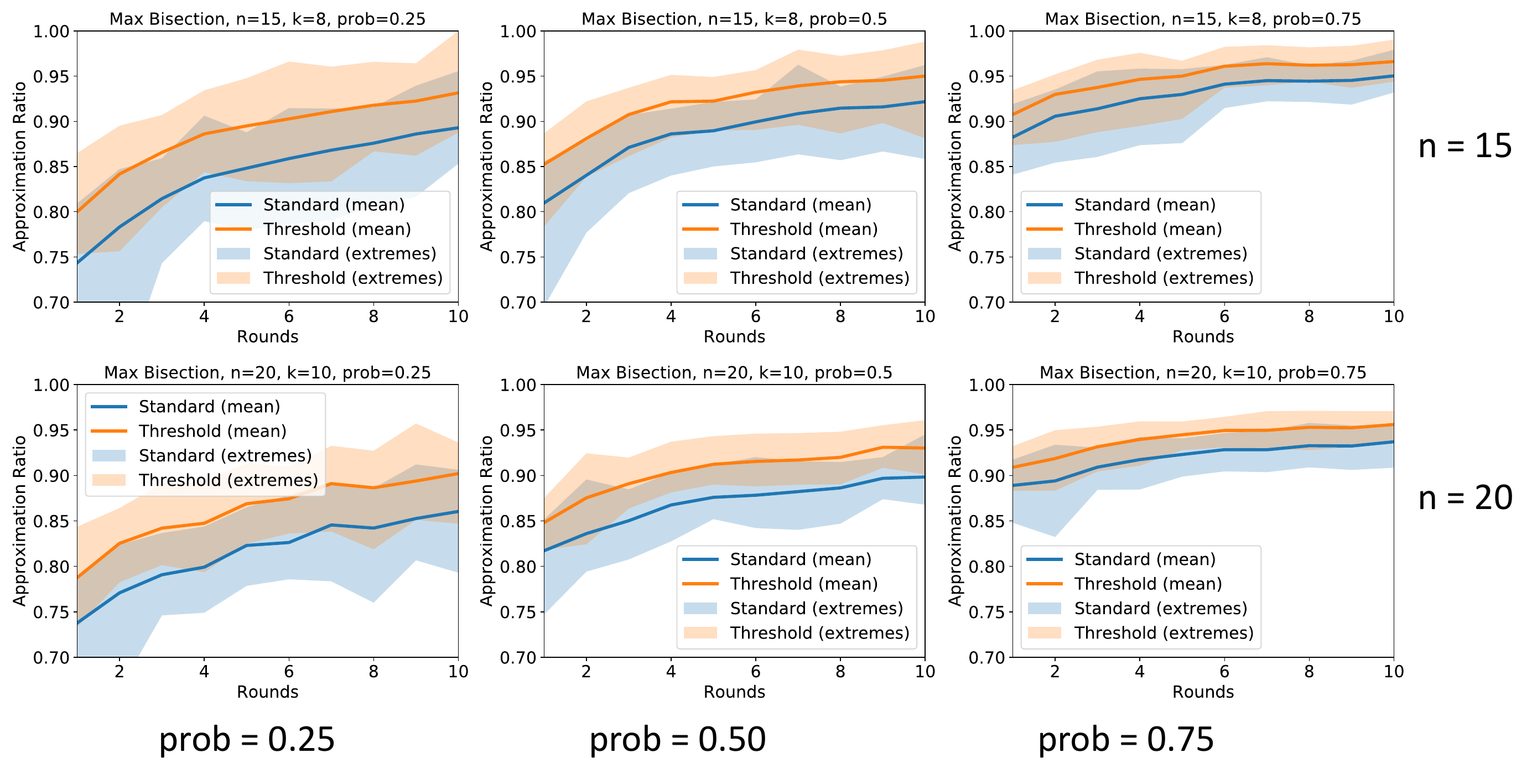}
\caption{{\bf Maximum Bisection}: QAOA performance for graphs of size 15 (top row) and size 20 (bottom row) with varying edge probabilities (prob = 0.25, 0.50, 0.75 from left to right)}
\label{fig:bisection}
\end{figure*}

\begin{figure*}
\centering
\includegraphics[width=0.85\linewidth]{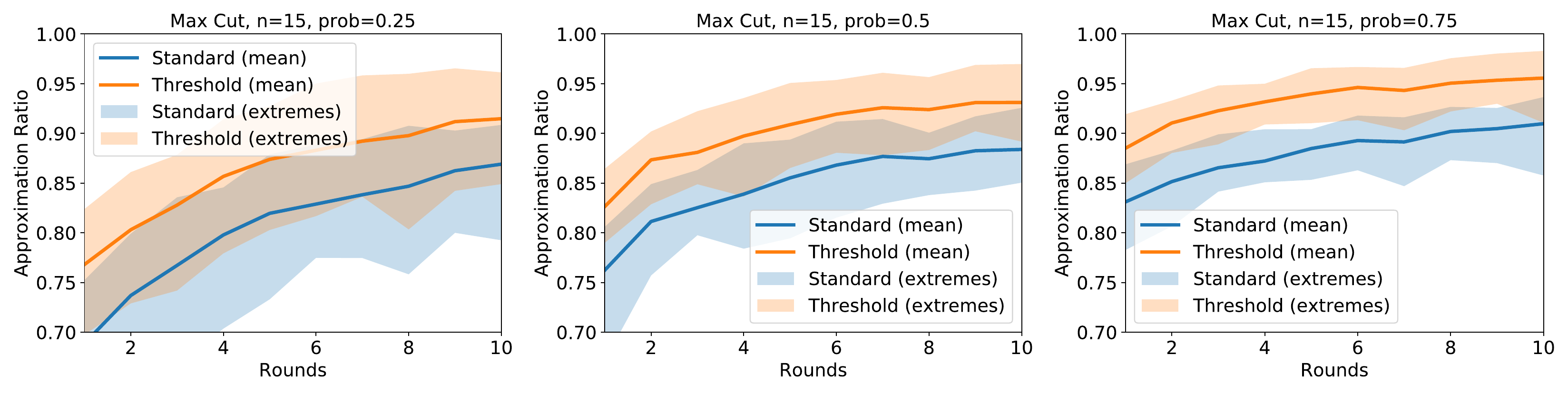}
\caption{{\bf Maximum Cut}: QAOA performance for different edge probabilites}
\label{fig:maxcut}
\end{figure*}

Our main result is that the Threshold-based GM-Th-QAOA finds better solutions as characterized by the approximation ratio across all optimization problems, graph sizes, edge probabilities, $k$ parameter values, and number of rounds. We did not expect such a clean sweep.

Figures \ref{fig:kds-probs}, \ref{fig:kds-ks}, \ref{fig:kvc-ks}, \ref{fig:bisection}, and \ref{fig:maxcut} show a select set of example plots for different values of edge probabilities, $k$ values and graph sizes for the four optimization problems. Each plot shows approximation ratio vs. the number of rounds for both the Threshold GM-Th-QAOA (red) and Standard GM-QAOA (blue), which means, of course, using the standard phase separator formulation as a Hamiltonian. The lighter shaded areas in red and blue show the minimum and maximum values found over the 30 random graphs tested. We use a different set of random graphs for each set of rounds in the plots, which explains the occasional surprising dip in performance at higher round counts. Both QAOA versions use the Grover mixer~\cite{baertschi2020grover}. 
The plots in these figures show a consistent outperformance of Threshold over Standard QAOA, albeit the relative improvement varies between less than one percent (for k-Vertex Cover with edge probability of $0.75$ and $k=15$) to around 8 percent (for k-Vertex Cover with edge probability of $0.25$ and $k=5$). The advantage of Threshold QAOA remains constant in almost all cases as we increase the number of rounds. These same general statements hold across all parameter combinations from the experimental design defined in Table \ref{choices}; we omit such additional plots as they look very similar to the selected plots presented here. 

We find this consistent outperformance to be somewhat astonishing. We emphasize that the outperformance is not due to the more efficient outer parameter search that GM-Th-QAOA allows for: all values on the plots are for the optimum outer parameter values. In fact, GM-Th-QAOA will not only find better solutions, but it will also find them faster because the outer loop parameter value finding is also much simpler for GM-Th-QAOA than for Standard GM-QAOA.

While most of these plots tell a very similar story and mainly serve to drive home the point of how our main finding is valid across a large number of parameter values, we still find a few interesting differences: Figure \ref{fig:kds-probs} shows our two QAOA versions for Maximum k-Densest Subgraph with increasing edge probabilities (0.25, 0.50, 0.75 from left to right) on graphs of size 20 with $k$-value set to 15. Threshold and Standard QAOA both improve their performance as edge probability is increased, and the relative performance difference actually becomes a little smaller as we increase edge probability. We also note that the curves flatten out at round count of 8 or higher. 

In contrast, Figure \ref{fig:kds-ks} shows the performance for $k= \{5, 10, 15\}$ again for graphs of size 20 and with edge probability fixed to $0.25$. Increasing $k$ again lead to improved performance and a narrower performance gap between the two QAOA versions.

Figure \ref{fig:kvc-ks} shows performance results for Maximum $k$-Vertex Cover of graphs of size 20 with edge probability of $0.75$ for different $k$-values. Threshold QAOA's performance advantage is less pronounced than it is for Maximum $k$-Densest Subgraph (note the y-axis values), but it still exists consistently. 

Figure \ref{fig:bisection}  studies the performance on the Maximum Bisection problem for graphs of size 15 (top row) and size 20 (bottom row) with increasing edge probabilities from left to right. The relative performance advantage of Threshold QAOA appears to be stable across the different graph sizes. 

Figure \ref{fig:maxcut} compares the QAOAs for Maximum Cut. Maximum Cut is the only unconstrained problem we considered, meaning all computational basis states are feasible solutions. Unlike other studies for MaxCut that limit node degrees to a small number, we use general random graphs. Thus the speed-up trick for constrained problems of only looking at feasible solutions no longer applies here. Simulation times thus start to get prohibitive a bit earlier here. Our plots show again that GM-Th-QAOA consistently outperforms GM-QAOA.

\subsection{GM-Th-QAOA Simulation Speed-up Tricks Allow Scaling to 100 qubits and 16,384 rounds}

Our tricks to speed up the numerical simulation of Threshold QAOA with the Grover Mixer enable us to classically simulate graph problems of 100 vertices, to near unlimited number of rounds. Such scaling greatly improves the confidence in our results, particularly with respect to ranking different QAOA variations and with respect to levels of approximation ratios achievable. Standard QAOA simulations so far (including our own) have mostly included simulations of graphs of up to 16 vertices and up to 10 rounds with performance and relative performance getting extrapolated from these small cases. 
A notable exception is a 25-round simulation of GM-QAOA for Maximum Satisfiability on 6 variables~\cite{akshay2020reachabilitydeficits}.

\begin{figure*}
\centering
\begin{subfigure}{.4\textwidth}
    \centering
    \includegraphics[width=\linewidth]{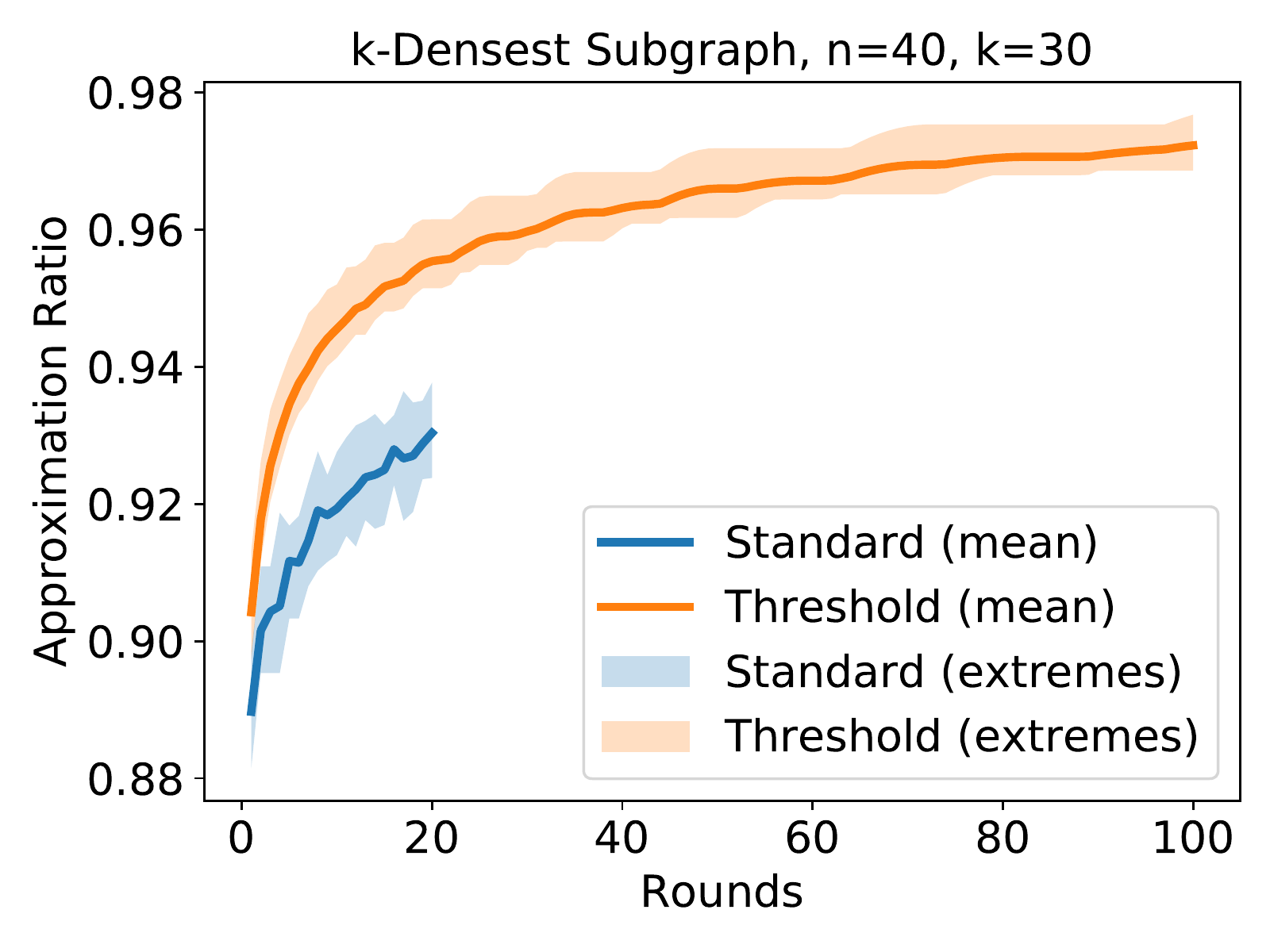}
\end{subfigure}
\hspace{0.05\linewidth}
\begin{subfigure}{.4\textwidth}
    \centering
    \includegraphics[width=\linewidth]{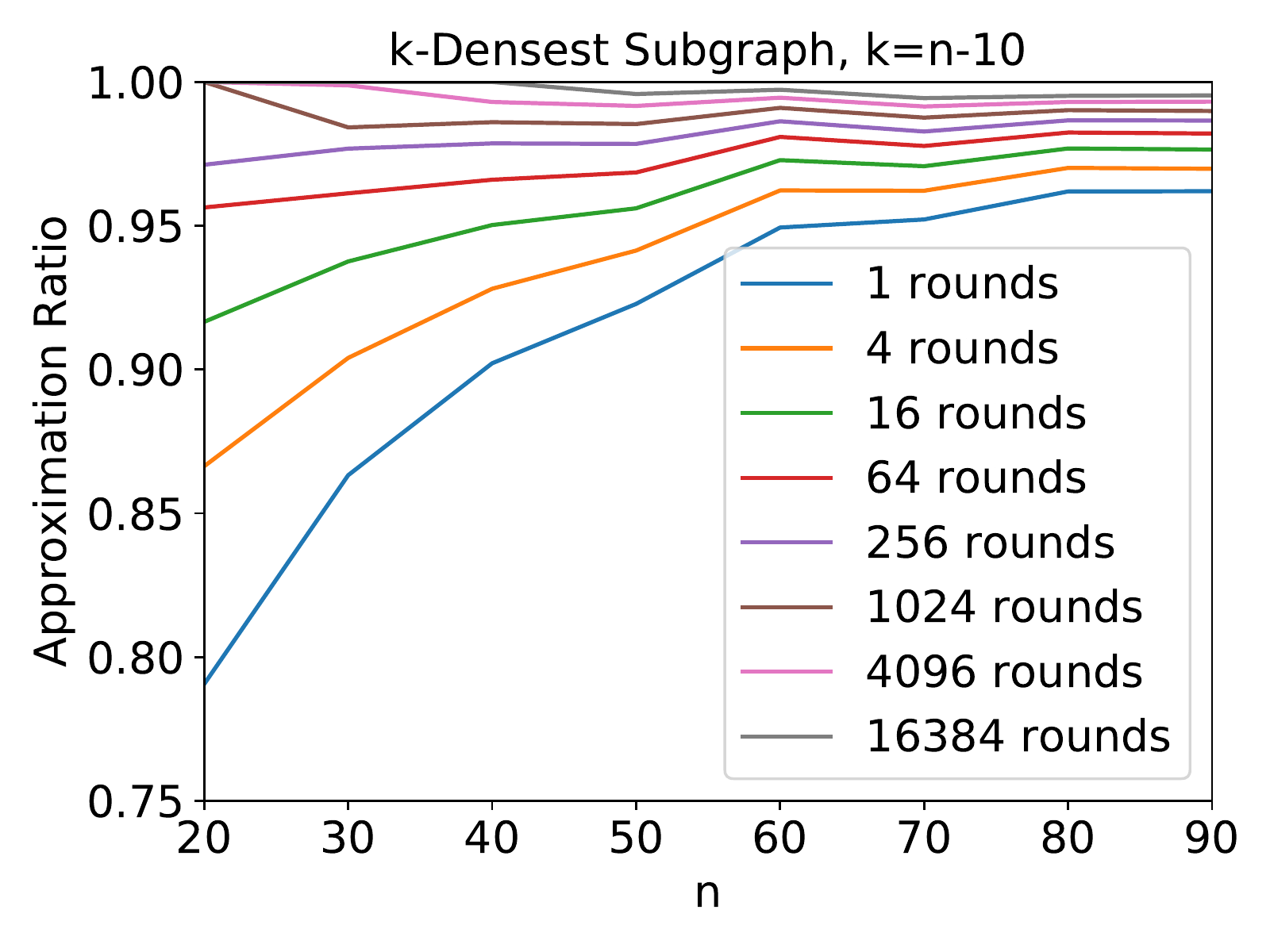}
\end{subfigure}
\caption{Approximation ratios for the $k$-Densest Subgraph problem at different scales: 
    (left) $k$-Densest Subgraph at intermediate scale of $n=40,\ k=30$. GM-Th-QAOA still outperforms standard GM-QAOA, but the parameter search of our simulation of GM-QAOA reaches its limits at 20 rounds.
    (right) Large-scale performance of GM-Th-QAOA for up to $n=100,\ p=2^{14}$.
}
\label{fig:scaling}
\end{figure*}

Figure \ref{fig:scaling} (left) shows the difference of the two QAOA versions for the $k$-Densest Subgraph problem on a medium-sized graph of 40 nodes with edge probability $0.5$. Clearly, Threshold QAOA outperforms Standard QAOA, and here the performance gap appears to even increase with the number of rounds. More importantly, however, classical simulation of Standard GM-QAOA becomes prohibitive on standard mid-range computing hardware, namely on the order of a few minutes per graph vs. milliseconds for GM-Th-QAOA. 

Figure \ref{fig:scaling} (right) only shows GM-Th-QAOA runs for a single graph at each graph size up to $n=100$ with edge probability $0.5$ and $k=n-10$. To show the effect of different number of rounds $p$ allowed, we draw lines for various values of $p$. Clearly, a high round count results in better approximation ratios. The perhaps surprising effect that approximation ratios achieved improve with graph size $n$ for smaller round counts can be explained through the edge density: a randomly selected subset of $n-10$ vertices of a large graph will be closer to the optimum $n-10$ subset than that of a smaller graph.

\subsection{GM-Th-QAOA and the Grover speed-up}
\label{sec:qaoa-grover-speedup}


Threshold QAOA is a general-purpose quantum optimization heuristic with a subroutine runtime linear in the number of rounds $p$, which is usually chosen to be small.
When combined with the Grover Mixer~\cite{baertschi2020grover,akshay2020reachabilitydeficits}, we have observed that GM-Th-QAOA
shares some traits with quantum search algorithms, such as sharing phase shift angles $\pi$ (i.e., phase inversions) with the Grover oracle and diffusion operators~\cite{G96}, deploying an exponential quantum search~\cite{boyer1998}~(Section~\ref{sec:var-grover-angles-algorithm}) 
or adapting a threshold value as in the minimum/maximum finding algorithm~\cite{DH96}~(Section~\ref{sec:var-grover-threshold-algorithm}).

In particular, for a very high number of rounds $p$ (compared to the set of feasible solutions $S$, e.g. $p \in \Omega(\sqrt{|S|})$),
the behaviour of our threshold and angle finding subroutines show a similar behaviour as the standard quantum minimum/maximum finding and exponential search algorithms.
We do not endorse turning round count $p$ into a variational parameter for GM-Th-QAOA. However, the thought experiment shows that  GM-Th-QAOA will never fare worse than the minimum/maximum finding variant of Grover's quantum search and thus also achieve at least a quadratic speed-up. 

Thus, to QAOA advocates, this connection can be viewed as a formal proof that QAOA achieves at least some speed-up in an asymptotic setting. At the same time, QAOA skeptics will point to their long-standing assertion that QAOA will not outperform Grover search in a worst-case scenario. We have not found a formal proof to refute the skeptics' claim, but rather we have designed a pragmatic QAOA heuristic that allows us to fix a (quantum) computational time budget and find the best possible solution within such a budget, which we believe will be of practical value once error-corrected and scalable quantum computing arrives. 

%

\section{Related and Future Work}
\label{sec:related-work}

Grover's quantum search algorithm~\cite{G96} was the first quantum algorithm to promise a quadratic speed-up over combinatorial \emph{search} problems. 
When it comes to combinatorial \emph{optimization} problems, we may be interested in either searching for the exact optimum, or approximating the problem with a close-to-optimal solution. The first approach exact optimization was a minimum finding algorithm~\cite{DH96} based on the quantum exponential search algorithm~\cite{boyer1998}, which itself is a generalization of Grover search to an unknown number of solutions.
A different proposal for exact optimization was given in the form of the adiabatic algorithm~\cite{Farhi2000}.

Farhi et.~al. later introduced the Quantum Approximate Optimization Algorithm~\cite{Farhi2014} which can be seen as a low-order Trotterization of the adiabatic algorithm. Surprisingly, the minimum finding algorithm has not seen such an adaption to approximate optimization. To the best of our knowledge, this work is the first to close this gap, in the sense of Section~\ref{sec:qaoa-grover-speedup}.

On the other hand, approaches using selective phase shift versions of Grover's oracle and diffusion operators have been discussed for a while and exclusively in the search setting; starting with Grover's fixed point quantum search with phase shifts of $\pm \pi/3$~\cite{grover2005fixed}, to more general angles which can recover the quadratic speed-up while preserving convergence to a fixed interval~\cite{yoder2014fixed}, to fully variational Grover search proposals which may give a moderate increase in success probability~\cite{biamonte2018variational} while still (only) reflecting a Grover scaling~\cite{akshay2020reachabilitydeficits}.
A combination of a selective phase shift oracle with the transverse field based $X$-mixer in the search setting has been shown been shown to achieve a quadratic speed-up as well~\cite{jiang2017tfgrover}, with success probability and number of rounds a constant factor away from the optimum query complexity of Grover search~\cite{zalka1999groveroptimal}.

Finally, the selective-phase shift Grover diffusion operator has been introduced to the Quantum Alternating Operator Ansatz as the Grover Mixer for both unconstrained~\cite{akshay2020reachabilitydeficits} and constrained~\cite{baertschi2021grover} approximate optimization problems 
(albeit only in conjunction with standard phase separation), where it has been shown to moderately outperform the transverse field based $X$-mixer on unconstrained~\cite{akshay2020reachabilitydeficits} and the $XY$-model Ring mixer on Hamming weight constrained problems~\cite{cook2020kVC,baertschi2020grover}.

For future work, it will be important to give a full comparison of both Standard and Threshold phase separators in combination with all known problem-specific mixers. However, new insights in order to scale up problem sizes for simulations of parameter combinations beyond GM-Th-QAOA will be a necessary prerequisite for such future studies.  

\bibliographystyle{plainurl}
\bibliography{references}

\end{document}